BOLETIN DE LA SOCIEDAD ESPAÑOLA DE
# Cerámica y Vidrio
A R T I C U L O# Microestructura y propiedades mecánicas a altas temperaturas del α-SiC sinterizado con fase líquida de $Y_2O_3$-$Al_2O_3$

M. CASTILLO-RODRÍGUEZ, A. MUÑOZ Y A. DOMÍNGUEZ-RODRÍGUEZ
Departamento de Física de la Materia Condensada. Universidad de Sevilla.
Apartado 1065. 41080-Sevilla. EspañaMuestras de α-SiC han sido sinterizadas con fase líquida (LPS) a 1950 °C, en atmósfera de argón y tiempos de procesado entre 1 y 7 horas. Mediante microscopia electrónica de barrido (SEM) se ha caracterizado la microestructura, obteniéndose un tamaño medio de grano que aumenta con el tiempo de procesado desde 0.64 a 1.61 µm.

Estas muestras han sido deformadas en compresión a carga constante, a temperaturas entre 1450 y 1625 °C, tensiones entre 25 y 450 MPa y velocidades de deformación entre $4.2 \cdot 10^{-8}$ y $1.5 \cdot 10^{-6}$ $s^{-1}$. Se han determinado los parámetros de fluencia, obteniéndose para n valores entre 2.4 ± 0.1 y 4.5 ± 0.2 y Q=680 ± 35 kJ.$mol^{-1}$ en muestras sinterizadas durante 1 hora, y n entre 1.2 ± 0.1 y 2.4 ± 0.1 y Q=710 ± 90 kJ.$mol^{-1}$ para muestras sinterizadas durante 7 horas. Estos resultados se han correlacionado con la microestructura y se han propuesto como posibles mecanismos de deformación el deslizamiento de frontera de grano (GBS) acomodado por difusión en volumen, y el movimiento de dislocaciones, actuando ambos mecanismos de forma simultánea.

*Palabras clave: α-SiC, propiedades mecánicas, microestructura, sinterización con fase líquida, fluencia.*

**Microstructure and high temperatures mechanical properties of liquid-phase-sintered α-SiC with $Y_2O_3$-$Al_2O_3$ additions**

Samples of α-SiC have been sintered with liquid phase (LPS) to 1950 °C, in atmosphere of argon and processing times between 1 and 7 hours. Using scanning electron microscopy (SEM) the microstructure of the samples has been characterized, being obtained average grain size that grows with the processing time from 0.64 to 1.61 µm.

These samples have been deformed in compression under constant load, to temperatures between 1450 and 1625 °C, stresses between 25 and 450 MPa and strain rate between $4.2 \cdot 10^{-8}$ and $1.5 \cdot 10^{-6}$ $s^{-1}$. The creep parameters have been determined, obtaining for n values between 2.4 ± 0.1 and 4.5 ± 0.2 and Q=680 ± 35 kJ.$mol^{-1}$ in samples sintered during 1 hour and n between 1.2 ± 0.1 and 2.4 ± 0.1 and Q=710 ± 90 kJ.$mol^{-1}$ for samples sintered during 7 hours. These results have been correlated with the microstructure, being the grain boundary sliding (GBS) accommodated by lattice diffusion and the movement of dislocations the possible deformation mechanisms, operating both in a simultaneous way.

*Keywords: α–SiC, mechanicals properties, microstructure, liquid phase sintered, creep.*## 1. INTRODUCCIÓN

El carburo de silicio SiC es un cerámico avanzado de gran interés para aplicaciones estructurales debido a su elevada resistencia a la corrosión, excelentes propiedades mecánica tanto a baja como a alta temperatura, buena conductividad térmica y bajo coeficiente de dilatación térmica. En gran parte de estas aplicaciones el material se encuentra sometido a elevadas tensiones de origen mecánico o térmico, como consecuencia de gradientes de temperatura, por lo que el conocimiento de las propiedades mecánicas del material a altas temperaturas juega un papel importante para muchas de estas aplicaciones.

En la bibliografía existen diversos estudios sobre las propiedades mecánicas del SiC a alta temperatura, fundamentalmente estudios de deformación a carga constante, sin embargo existe cierta dispersión en los resultados debido a que los mecanismos de deformación dependen en gran medida de la microestructura de las muestras y ésta es función fundamentalmente del método de obtención y también de otros parámetros como impurezas o aditivos de sinterización [1-9]. En el SiC policristalino el carácter covalente del enlace Si-C, el bajo coeficiente de difusión de las especies atómicas y la alta energía de las fronteras de grano, hacen necesario temperaturas y presiones de sinterizado muy elevadas. Como consecuencia de esta dificultad de procesado, distintas técnicas de fabricación del SiC policristalino se han desarrollado durante la última década, dando lugar a materiales con características muy diversas.

Lane y col. [1] estudian un α-SiC policristalino de 3.7 µm de tamaño medio de grano, obtenido por sinterización a 2100 °C con pequeñas cantidades de C (0.5 % en peso), B (0.42 % en peso) y otras impurezas en menores cantidades (por debajo de 100 p.p.m.). El mecanismo de deformación propuesto por estos autores es el deslizamiento de frontera de grano (GBS) acomodado por difusión en volumen, para temperaturas por debajo de 1650 °C. A temperaturas superiores encuentran que la actividad de dislocaciones contribuye de forma significativa a los mecanismos de deformación. Nixon y col. [2] amplían el estudio de Lane para diferentes tamaños de grano. Los resultados obtenidos son similares, encontrando que la mayor o menor contribución del movimiento de dislocaciones a los

278                                                                                                          Bol. Soc. Esp. Ceram. V., 44 [5] 278-285 (2005)

mecanismos de deformación es función del tamaño de grano.

Backhaus-Ricoult y col. [3] estudian las propiedades mecánicas a alta temperatura de dos tipos de SiC policristalino obtenidos por prensado isostático en caliente (HIP) a temperatura de 2000 ºC y presión de 2 kbar. Uno de ellos sin impurezas y otro con impurezas de B, Si, Fe, O y C (por debajo de 2000 p.p.m.). Estos autores proponen como mecanismo de deformación el GBS acomodado por difusión en la frontera de grano y cavitación, para tensiones de deformación inferiores a 500 MPa; a tensiones superiores la frontera de grano no supone un obstáculo para el movimiento de dislocaciones a través de ella por lo que el movimiento de dislocaciones es el mecanismo que controla la deformación. Por otra parte, encuentran un incremento en la velocidad de deformación como consecuencia de la presencia de los aditivos.

Jou y col. [4] obtienen SiC policristalino, de tamaño medio de grano 2 $\mu$m, sinterizado con fase líquida de Al$_2$OC (10 % en peso) a una temperatura de 1875 ºC. Como resultado de los ensayos de deformación a carga constante en este material los autores concluyen que la deformación a temperaturas entre 1500 y 1650 ºC es controlada simultáneamente por procesos de difusión en volumen y movimiento de dislocaciones.

Hamminger y col. [5] realizan ensayos de deformación a carga constante en diferentes SiC policristalinos dopados con aluminio y carbón y con boro y carbón y concluyen que, a temperaturas entre 1470 y 1660 ºC y tensiones entre 100 y 190 MPa, la deformación en estos materiales está controlada por procesos de difusión en volumen.

Gallardo-López y col. [6] estudian el SiC policristalino sinterizado con pequeñas cantidades de fase líquida de Y$_2$O$_3$ y Al$_2$O$_3$ (inferior al 2% en volumen) y 1.2 $\mu$m de tamaño de grano. El GBS acomodado por difusión en volumen y la actividad de dislocaciones, actuando de forma simultánea, son los mecanismos de deformación propuestos por estos autores para temperaturas entre 1575 y 1700 ºC y tensiones de deformación entre 90 y 500 MPa.

Para el SiC policristalino con silicio libre (entre 10 y 12 % en volumen), obtenido por reacción del Si en estado gaseoso con una preforma porosa de SiC que contiene carbono libre, Carter y col. [7] proponen la actividad dislocaciones como mecanismo que controla la deformación a temperaturas entre 1575 y 1650 ºC y tensiones comprendidas entre 110 y 220 MPa.

Muñoz y col. [8] y Martínez-Fernández y col. [9] han estudiado las propiedades mecánicas a alta temperatura del SiC policristalino obtenido por reacción (RFSC) con diferentes cantidades de Si remanente (9 y 20% en volumen) y con una fase remanente de Si + NbSi$_2$ en una proporción del 9% en volumen. El estudio se ha realizado mediante ensayos de compresión a velocidad constante y a carga constante. En estos materiales la deformación a temperaturas entre 1300 y 1350 ºC y carga constante tiene lugar por deformación de la fase intergranular remanente, de forma que la velocidad de deformación disminuye con la deformación hasta detenerse como consecuencia del contacto entre los granos de SiC. Los resultados obtenidos se explican mediante un modelo de fluencia con presencia de fase viscosa intergranular.

En este trabajo se estudia el comportamiento mecánico a alta temperatura de un material policristalino de SiC obtenido mediante sinterización con fase líquida (LPS) de Y$_2$O$_3$ y Al$_2$O$_3$, con una cantidad moderada de estos aditivos (10 % en peso). El material se ha sinterizado en atmósfera de argón durante diferentes tiempos (1, 3, 5 y 7 horas) con el fin de estudiar la influencia del tiempo de procesado en la microestructura del material. El efecto del tiempo de procesado en las propiedades mecánicas a alta temperatura se ha estudiado para las muestras sinterizadas durante 1 y 7 horas. Los resultados obtenidos son discutidos y comparados con los existentes en la bibliografía para diferentes SiC policristalinos.

## 2. MÉTODO EXPERIMENTAL

Los materiales estudiados en este trabajo se han obtenido por sinterización con fase líquida (LPS), a partir de polvo α-SiC y utilizando como aditivos Y$_2$O$_3$ y Al$_2$O$_3$ en la proporción 3/5 para obtener como fase intergranular el granate de itrio y aluminio Y$_3$Al$_5$O$_{12}$ (YAG). Las muestras están compuestas por un 90 % en peso de α-SiC y un 10 % en peso de YAG. El procesado se ha realizado a 1950 ºC, en atmósfera de argón durante diferentes tiempos (1, 3, 5 y 7 horas), obteniéndose una densificación próxima al 98 %. Estos materiales han sido fabricados en el Department of Metallurgy and Material Science Engineering, Institute of Materials Science, University of Connecticut, USA. Más detalles sobre el procesado pueden ser consultados en los trabajos de Padture y col. [10] y Pujar y col. [11].

El estudio microestructural de las muestras se ha realizado por microscopia electrónica de barrido (SEM), utilizando electrones secundarios y retrodispersados, en un microscopio electrónico de barrido Philips XL-30 con tensiones de trabajo de hasta 30 kV. Las muestras han sido previamente pulidas con papel abrasivo y pasta de diamante de tamaño de grano decreciente hasta 1 $\mu$m de diámetro. Posteriormente se ha revelado la frontera de grano mediante ataque con plasma de CF$_4$ mezclado con O$_2$ en la proporción 1/6, durante dos horas aproximadamente (Plasma Asher K1050X, Emitech). Por último, en algunos casos, ha sido necesario depositar sobre la superficie una lámina delgada de Au para favorecer la conducción evitando efectos de carga durante las observaciones microscópicas. Los parámetros morfológicos de los granos, para los distintos tipos de muestras, se han determinado analizando entre 350 y 450 granos, en micrografías de diferentes regiones, mediante un analizador de imágenes semiautomático (Videoplan MOP 30, Kontron Electronik).

Los ensayos mecánicos se han realizado con muestras sinterizadas durante 1 y 7 horas. Estas muestras, de dimensiones aproximadas 2.5 x 2.5 x 5 mm, han sido sometidas a ensayos de compresión a carga constante en atmósfera de argón. Se han efectuado en total 6 experiencias de fluencia, cada una de ellas con una duración aproximada de 350 horas. Los ensayos se han realizado en una máquina prototipo [12], a temperaturas entre 1550 y 1625 ºC y tensiones entre 25 y 450 MPa.

## 3. RESULTADOS Y DISCUSIÓN

### 3.1. Caracterización microestructural de las muestras

La necesidad de correlacionar la microestructura de los materiales utilizados en nuestro estudio con los resultados de los ensayos mecánicos a carga constante hace necesaria la caracterización inicial de los mismos en función del tiempo de procesado en atmósfera de argón.

Las imágenes obtenidas mediante SEM, a bajos aumentos (figura 1), muestran un material bifásico constituido por granos de SiC (contraste oscuro) y una fase secundaria (YAG) formada a partir de los aditivos de sinterización (contraste claro). Para los diferentes materiales estudiados se observa una distribución homogénea de la fase intergranular así como la presencia de cierta porosidad en todos ellos que se hace más significativa al ir aumentando el tiempo de sinterización.

Estas observaciones sobre la porosidad están de acuerdo con las medidas de la densidad realizadas por el método de Arquímedes, para los diferentes tipos de muestras. En la tabla I se indican los valores obtenidos y su comparación con la densidad teórica 3.310 ± 0.002 g.cm$^{-3}$. Se observa una disminución de la densidad al aumentar el tiempo de procesado en atmósfera de argón.

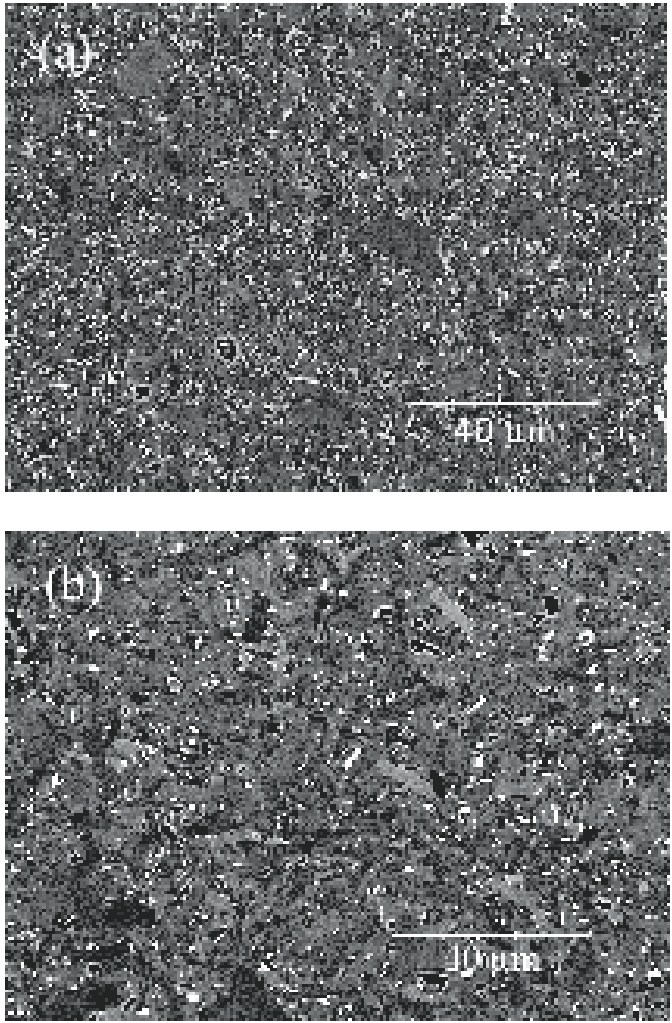

Fig. 1- Imágenes SEM de muestras sinterizadas en atmósfera de argón durante a) 1 hora y b) 7 horas. El contraste oscuro corresponde a los granos de SiC y el claro a la fase intergranular (YAG).

A mayores aumentos, la figura 2 muestra para los diferentes materiales una localización de la fase intergranular en los puntos triples de los granos de SiC y rellenando cavidades entre estos granos. En esta figura se observa también un aumento en el tamaño de grano al aumentar el tiempo de procesado. Este efecto se hace especialmente significativo para las muestras sinterizadas durante 7 horas, y va acompañado de un cambio importante en la forma de algunos granos que pasan de equiaxiados a muy alargados, con un incremento importante de tamaño. Esto da lugar a una distribución bimodal tanto en la forma como en el tamaño de los granos.

Igualmente se observa (figura 2) la subestructura núcleo-anillo en los granos de SiC, lo que indica que el crecimiento de grano tiene lugar mediante la disolución de los granos más pequeños en la fase líquida y la posterior precipitación de los átomos (C y Si) sobre las partículas de mayor tamaño. El núcleo corresponde a la partícula del polvo de partida (contraste más oscuro en el interior del grano) y el anillo al nuevo material depositado [13]. La diferencia de contraste es consecuencia de que el anillo presenta trazas de Y, Al y O procedentes de la fase líquida. En el caso de granos de tamaño elevado, esta subestructura es más difícil de observar ya que la partícula de partida es pequeña frente al tamaño del grano y por tanto difícil de interceptar al seccionar el grano.

El crecimiento anisótropo de algunos granos para tiempos de sinterizado de 7 horas puede justificarse por la presencia en el polvo α-SiC, utilizado en la sinterización, de cierta cantidad de fase β. En este caso, a la temperatura de procesado se produce la transformación β→α en el interior de estos granos, formando granos SiC bifásicos β/α de forma que los átomos (Si y C) disueltos tienden a precipitar como fase α sobre zonas α del grano (minimiza la energía) y esto hace que los granos bifásicos crezcan de manera altamente anisótropa, dando lugar a granos alargados. Este proceso de transformación del β-SiC a la temperatura de procesado está ampliamente documentado en la bibliografía [13-17] y es utilizado con cierta frecuencia para obtener SiC con una microestructura predeterminada. Estos trabajos muestran como la cinética de transformación está fuertemente influenciada por la densidad de faltas de apilamiento en los granos de β-SiC, por lo que es previsible que la transformación no sea completa ni aún en el caso de las muestras procesadas durante 7 horas.

El elevado tamaño de algunos granos junto a la forma alargada de los mismos dificulta la necesaria redistribución de la fase intergranular para que esta ocupe las cavidades que van quedando entre los granos, y ello hace que la porosidad de las muestras con tiempos de sinterizado de 7 horas sea superior al resto de las muestras con las que hemos trabajado. Efectivamente, para este tipo de muestras, se observa (figura 2d) que la mayor parte de la porosidad se encuentra localizada en las proximidades de granos alargados de elevado tamaño.

La figura 3 muestra los histogramas de distribución de tamaño de grano (diámetro planar equivalente, d=(4·área/π)$^{1/2}$ ) para las muestras sinterizadas durante 1 y 7 horas. Se observa, en el caso de las muestras sinterizadas durante 7 horas, un desplazamiento de la distribución hacia mayores tamaños de grano y un ensanchamiento de la misma tendiendo a una distribución bimodal debido a la presencia de los granos alargados de elevado tamaño.

Otros parámetros morfológicos analizados han sido el factor de forma de los granos, F=4π·área/perímetro$^2$, y el factor de aspecto $F_{asp}$, cociente entre la longitud máxima y la longitud mínima perpendicular a ésta, en la sección planar de los granos. En la tabla II se presentan los valores medios obtenidos para el tamaño de grano así como para el resto de parámetros morfológicos medidos. En ella se observa el crecimiento de grano que tiene lugar al aumentar el tiempo de procesado y el alejamiento de la forma equiaxiada de los granos para tiempos de procesado de 7 horas. En la determinación de las incertidumbres tanto de los valores medios como de sus desviaciones estándar se ha considerado que en primera aproximación estas magnitudes presentan una distribución gaussiana.

TABLA I. VALORES DE LA DENSIDAD DE LAS MUESTRAS Y SU COMPARACIÓN CON LA DENSIDAD TEÓRICA.

| Tiempo de procesado (h) | 1 | 3 | 5 | 7 |
|---|---|---|---|---|
| Densidad medida (g/cm³) | 3.27 ± 0.02 | 3.22 ± 0.03 | 3.20 ± 0.02 | 3.14 ± 0.03 |
| Densidad medida / densidad teórica (%) | 98.8±0.6 | 97.3 ± 1.0 | 96.7 ± 0.6 | 94.9 ± 1.0 |

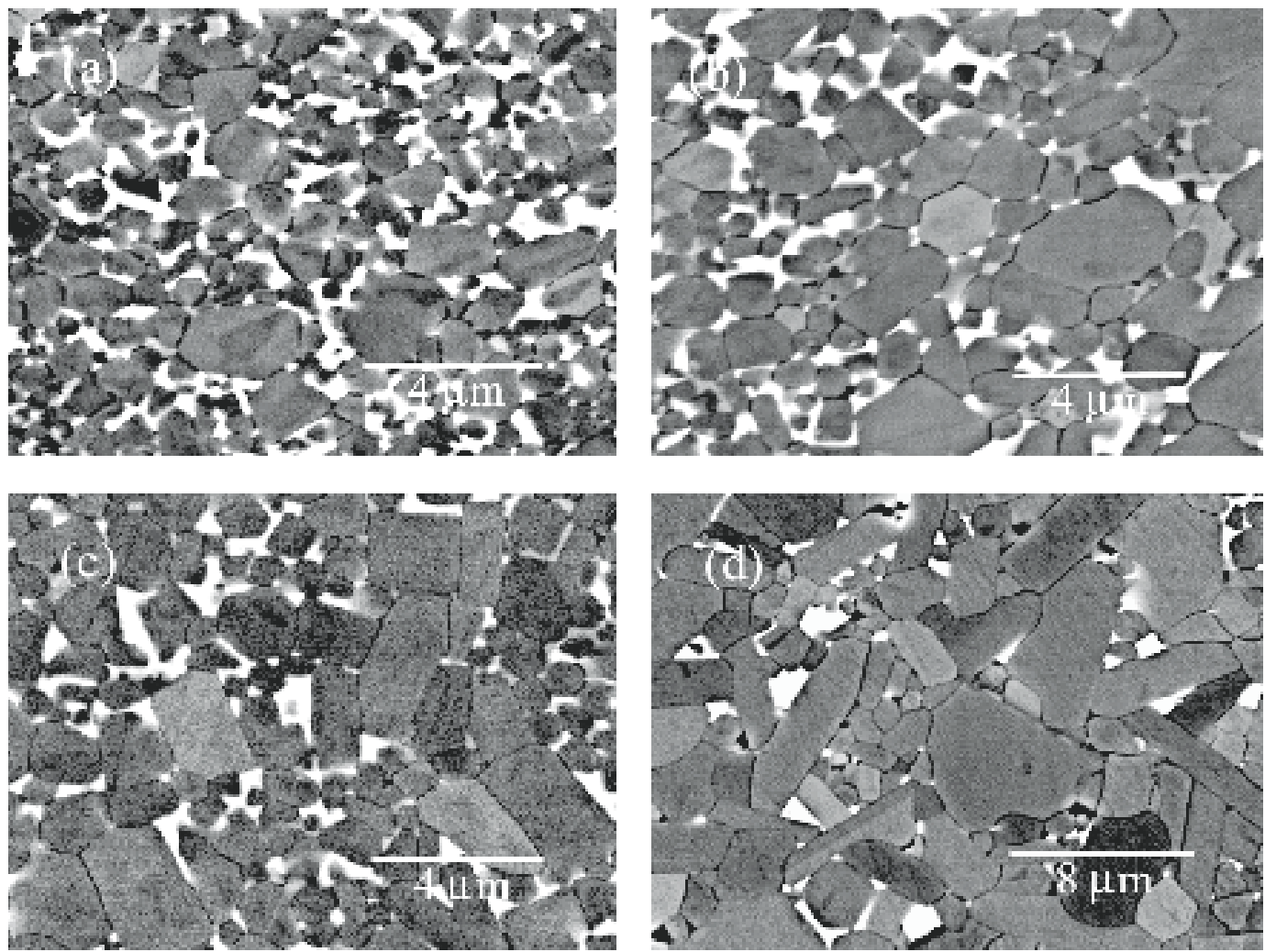

Fig. 2- Imágenes SEM de muestras sinterizadas en atmósfera de argón durante a) 1 hora, b) 3 horas, c) 5 horas y d) 7 horas. Las zonas oscuras corresponden a los granos de SiC y las claras a la fase intergranular (YAG).

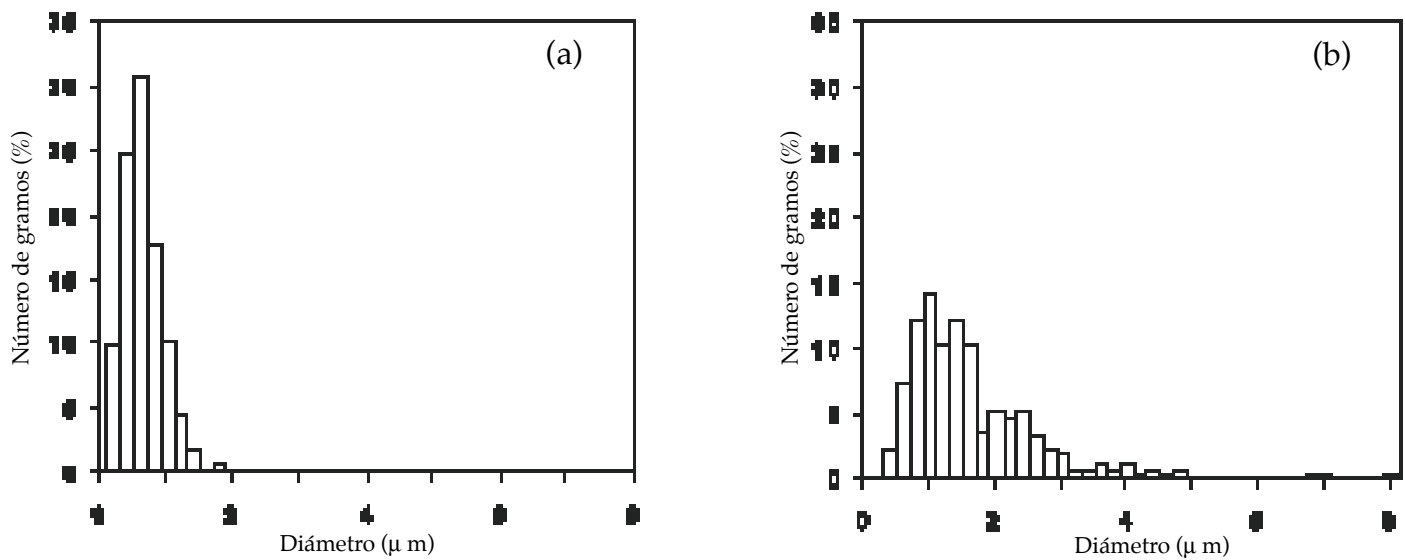

Fig. 3- Distribución del tamaño de grano en muestras sinterizadas en atmósfera de argón para diferentes tiempos de procesado. a) 1 hora, b) 7 horas.

### 3.2. Ensayos de compresión a carga constante

Los ensayos de deformación a carga constante (fluencia) se han realizado para muestras sinterizadas durante 1 y 7 horas. Para el análisis de los mismos se ha utilizado la ecuación fenomenológica:

$$\dot{\varepsilon} = A \frac{\sigma^n}{d^p} \exp\left(-\frac{Q}{kT}\right) \quad [1]$$

donde A es un parámetro que recoge la dependencia de la velocidad de deformación con las características microestructurales del material (morfología de los granos, presencia y distribución de fases secundarias, porosidad,...), σ la tensión aplicada sobre la muestra, d el diámetro medio de los granos, k la constante de Boltzmann, T la temperatura absoluta de la experiencia, n y p los exponentes de tensión y tamaño de grano y Q la energía de activación del proceso responsable de la deformación. El valor de los parámetros n, p y Q constituye una información importante en la determinación de los mecanismos responsables de la deformación y su valor se obtiene de los ensayos de fluencia.

La figura 4 muestra $\dot{\varepsilon}$ en función de $\varepsilon$ para una muestra sinterizada durante 1 hora y deformada entre 1600 y 1550 ºC. En ella puede verse que después de un régimen transitorio inicial, en el que la velocidad de deformación cambia rápidamente, se alcanza un régimen estacionario, caracterizado por una velocidad de deformación aproximadamente constante, función de la temperatura y de la tensión aplicada. Como se observa en la figura, la modificación de alguna de estas variables, temperatura T o tensión aplicada σ, conduce a un nuevo estado estacionario. El valor de n o de Q se determina a partir del cambio en la velocidad estacionaria de deformación al modificar durante el ensayo el parámetro correspondiente σ o T, y utilizando la ecuación [1]. En la figura se incluyen los valores obtenidos de n y Q en los diferentes saltos de tensión y temperatura.

La tabla III resume los parámetros de ensayo y los valores medidos para los parámetros de fluencia n y Q en los dos tipos de muestras estudiadas. Los valores de n y Q se han calculado como promedio de los obtenidos en los diferentes saltos de tensión, para cada temperatura de ensayo, y en los diferentes saltos de temperatura, respectivamente.

Los valores medidos de la energía de activación concuerdan con gran parte de los valores obtenidos en trabajos previos existentes en la bibliografía, referidos a materiales policristalinos de SiC sinterizados por diferentes técnicas. Así, Lane y col. [1] obtienen valores de la energía de activación entre 338 y 434 kJ.mol$^{-1}$, para temperaturas de ensayo por debajo de 1650 ºC y entre 802 y 914 kJ.mol$^{-1}$ para mayores temperaturas. Nixon y col. [2] encuentran valores de la energía de activación entre 387 y 541 kJ.mol$^{-1}$ para T<1650 ºC y entre 838 y 877 kJ.mol$^{-1}$ para T>1650 ºC. Backhaus-Ricoult y col. [3] obtienen una energía de activación de 364 kJ.mol$^{-1}$ en materiales sin impurezas y de 453 kJ.mol$^{-1}$ en materiales con impurezas de B, Si, Fe, O y C, para tensiones inferiores a 500 MPa y un valor de Q = 629 kJ.mol$^{-1}$ para tensiones por encima de 500 MPa. Igualmente, Jou y col. [4] encuentran un valor de la energía de activación de 743 kJ.mol$^{-1}$ para temperaturas de ensayo entre 1500 y 1650 ºC, Hamminger y col. [5] obtienen un valor de Q= 796 kJ.mol$^{-1}$, Gallardo-López y col. [6] obtienen un valor de Q = 840 kJ.mol$^{-1}$ a temperaturas entre 1575 y 1700 ºC y Carter y col. [7] encuentran un valor de Q = 711 kJ.mol$^{-1}$ entre 1575 y 1700 ºC.

Por otro lado, los valores de Q obtenidos en nuestro trabajo están de acuerdo con los valores que se encuentran en la bibliografía para la energía de activación medida por difusión en volumen del carbono (715 – 840 kJ.mol$^{-1}$) [18, 19] y del silicio (695 – 910 kJ.mol$^{-1}$) [20, 21] en el SiC, lo que sugiere que los mecanismos responsables de la fluencia en este material están controlados por procesos de difusión en el interior de los granos. Dado que el coeficiente de difusión del Si es, al menos, un orden de magnitud inferior al del C en el rango de temperaturas de nuestras experiencias, debe ser la difusión en volumen del Si el proceso que controla la velocidad de deformación. Los datos anteriores de la difusión en el SiC han sido medidos en α–SiC monocristalino, puro y dopado con nitrógeno (620 p.p.m.), y en β–SiC policristalino dopado con Cu (60 p.p.m.).

En relación con los valores obtenidos en nuestro trabajo para el exponente de tensión, n = 2.4 ± 0.1 y n = 4.5 ± 0.2 en muestras sinterizadas durante 1 hora y n = 1.2 ± 0.1 y n = 2.4 ±0.1 en muestras sinterizadas durante 7 horas, cabe indicar

TABLA II. VALORES DE LOS PARÁMETROS MORFOLÓGICOS MEDIDOS ANTES DE LOS ENSAYOS MECÁNICOS, PARA LOS DIFERENTES TIPOS DE MUESTRAS

| Tiempo de procesado(h) | Diámetro (μm) | | Factor de forma | | Factor de aspecto | |
|---|---|---|---|---|---|---|
| | $\overline{d}$ | $\sigma_d$ | $\overline{F}$ | $\sigma_F$ | $\overline{F}_{asp}$ | $\sigma_{Fasp}$ |
| 1 | 0.64 ± 0.02 | 0.29 ± 0.01 | 0.82 ± 0.01 | 0.11 ± 0.01 | 1.52 ± 0.02 | 0.34 ± 0.01 |
| 3 | 0.97 ± 0.04 | 0.82 ± 0.05 | 0.85 ± 0.01 | 0.09 ± 0.01 | 1.48 ± 0.02 | 0.37 ± 0.01 |
| 5 | 1.21 ± 0.05 | 0.73 ± 0.03 | 0.85 ± 0.01 | 0.09 ± 0.01 | 1.54 ± 0.02 | 0.39 ± 0.01 |
| 7 | 1.61 ± 0.05 | 0.99 ± 0.06 | 0.78 ± 0.01 | 0.12 ± 0.01 | 1.87 ± 0.04 | 0.66 ± 0.04 |

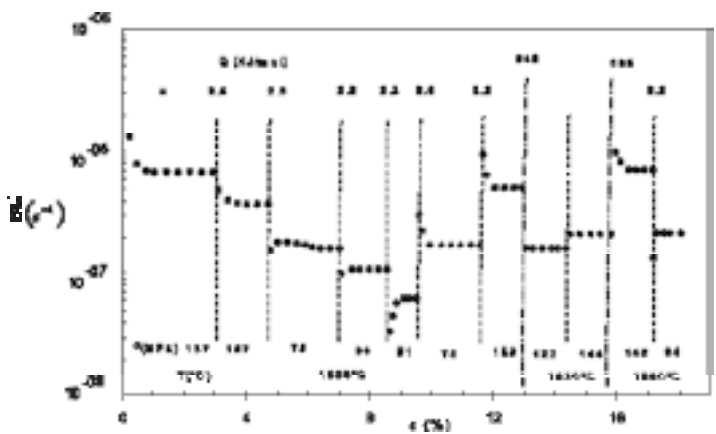

Fig. 4- Velocidad de deformación $\dot{\varepsilon}$ frente a la deformación $\varepsilon$ en el ensayo de deformación a carga constante de una muestra sinterizada en atmósfera de argón durante 1 hora. Las temperaturas de trabajo han sido 1600 y 1550 ºC y las tensiones aplicada entre 50 y 150 MPa. Se incluyen los valores calculados para n y Q en los diferentes saltos de tensión y temperatura.

bibliografía [1-7] para explicar la fluencia del SiC obtenido por diferentes técnicas.

Sin embargo los valores obtenidos del exponente de tensión n son superiores a la unidad que es el valor que corresponde al modelo clásico de GBS acomodado por difusión en volumen [22] aunque, como hemos indicado anteriormente, están dentro del rango de valores experimentales encontrados en la bibliografía en trabajos previos sobre el SiC. Estos valores de n superiores a la unidad sugieren que junto a la difusión en volumen debe existir otro mecanismo con igual energía de activación y mayor exponente de tensión, de forma que ambos mecanismos contribuyen de manera independiente a la deformación o a la acomodación del deslizamiento de frontera de grano durante la deformación, permitiendo alcanzar valores de deformación relativamente altos (superiores al 20 %).

El hecho de que el exponente de tensión crece al aumentar la tensión a que se somete la muestra sugiere la posibilidad de que el movimiento de dislocaciones en el interior de los granos sea el segundo mecanismo responsable de la acomodación del GBS o incluso sea el mecanismo primario de deformación que también son comparables con gran parte de los resultados recogidos en la bibliografía para el SiC policristalino fabricado por diferentes técnicas. Así, Lane y col. [1] obtienen valores que aumentan con la temperatura desde 1.44 a 1.71, Nixon y col. [2] obtienen valores entre 1.0 y 1.4 para T<1650 ºC y entre 1.4 y 2.5 para T>1650 ºC, Jou y col. [4] encuentran un valor de 1.74, Hamminger y col. [5] encuentran valores próximos a 1, Gallardo-López y col. [6] obtienen el valor n = 1.6, Carter y col. [7] obtienen n = 5.7 y Backhaus–Ricoult y col. [3] encuentran para bajas tensiones (inferiores a 500 MPa) el valor n = 1.5 y, para tensiones superiores, valores entre 3.5 y 4.

TABLA III. VALORES DE LOS PARÁMETROS DE ENSAYO Y DE LOS PARÁMETROS DE FLUENCIA n Y Q OBTENIDOS DE LAS DIFERENTES EXPERIENCIAS DE COMPRESIÓN A CARGA CONSTANTE.

| | Temperatura (ºC) | 1625 | 1600 | 1580 | 1550 |
|---|---|---|---|---|---|
| **Tiempo de procesado 1 hora** | σ (MPa) | 26 - 63 | 51 - 142 | | 200 - 442 |
| | $\dot{\varepsilon}$ (s$^{-1}$) | 9.6·10$^{-8}$–7.4·10$^{-7}$ | 6.3·10$^{-8}$–7.1·10$^{-7}$ | | 4.2·10$^{-8}$–1.5·10$^{-6}$ |
| | n | 2.4 ± 0.1 | 2.4 ± 0.1 | | 4.5 ± 0.2 |
| | Q | 685 ± 35 kJ·mol$^{-1}$ | | | |
| **Tiempo de procesado 7 horas** | σ (MPa) | 66 - 204 | 157 - 318 | 211 - 384 | |
| | $\dot{\varepsilon}$ (s$^{-1}$) | 5.8·10$^{-8}$–2.3·10$^{-7}$ | 4.3·10$^{-8}$–3.2·10$^{-7}$ | 9.1·10$^{-8}$–2.7·10$^{-7}$ | |
| | n | 1.2 ± 0.1 | 2.4 ± 0.1 | 2.4 ± 0.2 | |
| | Q | 710 ± 90 kJ·mol$^{-1}$ | | | |

Los parámetros morfológicos obtenidos de las observaciones microscópicas (SEM) realizadas sobre las muestras deformadas indican que no se han producido durante la fluencia cambios significativos en la forma y tamaño de los granos. La tabla IV muestra los valores de los parámetros morfológicos para los materiales sinterizados durante 1 y 7 horas y deformados a la más alta temperatura (1625 ºC). Estos valores son muy próximos a los obtenidos para los mismos tipos de muestras sin deformar (tabla II) lo cual indica que la deformación ha tenido lugar principalmente por cambio de posición de los granos en el interior de la muestra, es decir, por GBS. El valor obtenido para la energía de activación sugiere la difusión en volumen del Si como proceso de acomodación del deslizamiento de frontera de grano. Este mecanismo de deformación ha sido propuesto de forma general en la

(fluencia-restauración) de forma que la proporción en que este mecanismo contribuye al proceso global de deformación o acomodación es tanto mayor cuanto más elevada es la tensión aplicada sobre la muestra y más baja la temperatura.

Esta hipótesis ha sido verificada por otros autores. Así, Lane y col. [1] observaron por microscopia electrónica de transmisión (TEM) una elevada densidad de dislocaciones distribuidas homogéneamente en el volumen del grano. Sus micrografías muestran gran cantidad de faltas de apilamiento, bandas de deslizamiento y movimientos de subida de dislocaciones, por lo que el valor entre 1.44 y 1.71 que encuentran para n a temperaturas superiores a 1650 ºC lo justifican considerando el movimiento de dislocaciones como mecanismo de deformación que actúa de forma simultánea e independiente al deslizamiento de frontera de grano acomodado por procesos de difusión en volumen. Igual justificación dan Nixon y col. [2] a los exponentes de tensión superiores a la unidad (entre 1.4 y 2.5) encontrados a T>1650 ºC y Jou y col. [4] al valor n = 1.74 medido para tensiones entre 40 y 200 MPa. Backhaus-Ricoult y col. [3] también realizan un estudio por TEM de las muestras deformadas a altas tensiones, por encima de 500 MPa, observando una elevada densidad de dislocaciones. Estos autores consideran que a altos valores de la tensión la frontera de grano no es obstáculo para el movimiento de dislocaciones a través de ella, siendo este movimiento el mecanismo que de forma dominante controla la deformación para altas tensiones y esto justifica que el exponente de tensión alcance los valores entre 3.5 y 4 que ellos encuentran.

TABLA IV: VALORES DE LOS PARÁMETROS MORFOLÓGICOS DE MUESTRAS SOMETIDAS A ENSAYOS DE FLUENCIA A 1625 ºC.

| Tiempo de procesado (h) | Diámetro (µm) | | Factor de forma | | Factor de aspecto | |
|---|---|---|---|---|---|---|
| | $\overline{d}$ | $\sigma_d$ | $\overline{F}$ | $\sigma_F$ | $\overline{F}_{asp}$ | $\sigma_{Fasp}$ |
| 1 | 0.72 ± 0.02 | 0.35 ± 0.01 | 0.82 ± 0.01 | 0.09 ± 0.01 | 1.57 ± 0.02 | 0.39 ± 0.01 |
| 7 | 1.68 ± 0.05 | 1.06 ± 0.07 | 0.80 ± 0.01 | 0.11 ± 0.01 | 1.71 ± 0.02 | 0.40 ± 0.01 |

Por ultimo, indicar que también en nuestro grupo Gallardo-López y col. [6] en un trabajo previo sobre el SiC sinterizado con fase líquida, con una cantidad más pequeña de fase intergranular (inferior al 2% en volumen), han realizado un estudio mediante TEM del material deformado. Las micrografías muestran también una elevada densidad de dislocaciones en el volumen del grano, observándose faltas de apilamiento, bandas de deslizamiento y movimientos de subida de dislocaciones entre planos de deslizamiento paralelos. Este movimiento de dislocaciones permitió justificar el valor n=1.6 medido a tensiones entre 90 y 500 MPa.

Un estudio por TEM de la microestructura de dislocaciones está en curso para verificar en nuestras muestras la contribución de la actividad de dislocaciones a la deformación del material. De acuerdo con los resultados de los ensayos mecánicos esta contribución debe ser mayor en las muestras sinterizadas durante una hora que en las sinterizadas durante 7 horas, posiblemente por su diferente microestructura.

Las velocidades de deformación $\dot{\varepsilon}$ obtenidas en nuestros ensayos se encuentran entre $4.2 \cdot 10^{-8}$ y $1.5 \cdot 10^{-6}$ s$^{-1}$ para muestras sinterizadas durante 1 hora y entre $4.3 \cdot 10^{-8}$ y $3.2 \cdot 10^{-7}$ s$^{-1}$ para muestras sinterizadas durante 7 horas. Para un mismo valor de la temperatura y tensión aplicada se observa una velocidad de deformación al menos un orden de magnitud inferior en el caso de las muestras sinterizadas durante 7 horas. Esta diferencia es consecuencia de la distinta microestructura que presentan ambos tipos de muestras. La microestructura de granos alargados y entrecruzados que presentan las muestras sinterizadas durante 7 horas y su mayor tamaño, d=1.6 µm, en comparación con la microestructura de granos equiaxiados de menor tamaño, d=0.6 µm, de las muestras sinterizadas durante 1 hora (figura 2) justifica su diferente velocidad de deformación durante los ensayos de fluencia.

## 4. CONCLUSIONES

La caracterización microestructural del material antes de los ensayos mecánicos muestra una distribución homogénea de fase intergranular y una densificación próxima al 99 % para muestras sinterizadas durante 1 hora, que disminuye al 95 % al aumentar el tiempo de procesado hasta 7 horas. El tamaño medio de grano crece desde 0.64 a 1.61 µm al aumentar el tiempo de procesado de 1 a 7 horas, produciéndose un alejamiento de la forma equiaxiada en parte de los granos. Este efecto es especialmente significativo para 7 horas de sinterización. La morfología de los granos no se altera de forma apreciable durante la deformación.

Los valores de los parámetros de fluencia y los resultados del estudio microestructural sugieren que la deformación se produce por GBS acomodado por la difusión en volumen del Si y por activación de dislocaciones, actuando ambos mecanismos de forma simultánea. La contribución de la actividad de dislocaciones al proceso global es tanto mayor cuanto más elevada es la tensión aplicada y menor la temperatura.

Las muestras sinterizadas durante 7 horas son más resistentes a la deformación, de forma que, para unas mismas condiciones de ensayo, su velocidad de deformación es un orden de magnitud inferior a la de las sinterizadas durante 1 hora.

## REFERENCIAS

1. J. E. Lane, C. H. Carter, Jr. y R. F. Davis. "Kinetics and Mechanisms of high-temperature creep in silicon carbide: II, sintered α–SiC". J. Am. Ceram. Soc., 71 (4), 281-295 (1988).
2. R. D. Nixon y R. F. Davis. "Diffusion accommodated grain boundary sliding and dislocation glide in the creep of sintered alpha silicon carbide". J. Am. Ceram. Soc., 75 (7), 1786-1795 (1992).
3. M. Backhaus-Ricoult, N. Mozdzierz y P. Eveno. "Impurities in silicon carbide ceramics and their role during high temperature creep". J. Phys. III. France, 3, 2189-2210 (1993).
4. Z.C. Jou y A.V. Virkar. "High temperature creep of SiC densified using a transient liquid phase". J. Mater. Res., 6 (9), 1945-1949 (1991).
5. R. Hammiger, G. Grathwoht y F.Thuemmler. "Microanalytical investigation of sintered SiC Part I: Bulk material and inclusions". J. Mater. Sci., 18, 353 (1983).
6. A. Gallardo-López, A. Muñoz, J. Martínez-Fernández y A. Domínguez-Rodríguez. "High temperature compressive creep of liquid phase sintered silicon carbide". Acta Mater., 47 (7), 2185-2195 (1999).
7. C. H. Carter, Jr., R.F. Davis y J. Bentley. "Kinetics and Mechanisms of high-temperature creep in silicon carbide: I, Reaction bounded". J. Am. Ceram. Soc 67 (10), 409-417 (1984).
8. A. Muñoz, J. Martínez-Fernández, A. Domínguez-Rodríguez y M. Singh. "High-temperature Compressive Strength of Reaction-formed Silicon Carbide (RFSC) Ceramics". J. Eur. Ceram. Soc., 18, 65-68 (1998).
9. J. Martínez Fernández, A. Muñoz, A. R. Arellano López, F. M. Valera Feria, A. Domínguez Rodríguez y M. Singh. "Microstructure-mechanical properties correlation in siliconized silicon carbide ceramics". Acta Mater., 51, 3259-3275 (2003).
10. N. P. Padture. "In situ-toughened silicon carbide". J. Am. Ceram. Soc., 77(2), 519-523 (1994).
11. V. V. Pujar, R. P. Jensen y N. P. Padture. "Densification of liquid-phase-sintered silicon carbide". J. Mater. Sci. Lett., 19[11], 1011-1014 (2000).
12. H. Gervais, B. Pellissier y J. Castaing. "Machine de fluage pour essais en compression à hautes températures de matériaux céramiques". Rev. Int. Htes Temp. et Refract., 15, 43-47 (1978).
13. A.L. Ortiz. "Control microestructural de cerámicos avanzados SiC sinterizados con fase líquida $Y_2O_3$ - $Al_2O_3$". Tesis Doctoral, Universidad de Extremadura (2002).
14. S. K. Lee y C. H. Lee. "Effects of α-SiC versus β-SiC starting powders on microstructure and fracture –toughness of SiC sintered with $Al_2O_3$-$Y_2O_3$ additives". J. Am. Ceram. Soc., 77, 1655-1658 (1994).
15. M. Nader, F. Aldinger y M. J. Hoffmann. "Influence of the α/β-SiC phase transformation on microstructural development and mechanical properties of liquid phase sintered silicon carbide". J. Mater. Sci., 34, 1197-1204 (1999).
16. Y. W. Kim, M. Mitomo y G. D. Zhan. "Mechanism of grain growth in liquid-phase-sintered β-SiC". J. Mater. Res., 14, 4291-4293 (1999).
17. H. Xu, T. Bhatia, S. A. Deshpande, N. P. Padture, A. L. Ortiz y F. L. Cumbrera. "Microstructural evolution in liquid-phase-sintered SiC: I, effect of starting SiC powder". J. Am. Ceram. Soc., 84, 1578-1584 (2001).
18. M. H. Hon y R. F. Davis. "Self-diffusion of C-14 in polycrystalline β-SiC". J. Mater. Sci., 14, 2411-2421 (1979).

19. M. H. Hon y R. F. Davis. "Self-diffusion of Carbon-14 in high purity and N-doped α-SiC single crystals". J. Am. Ceram. Soc., 63, 546-552 (1980).
20. M. H. Hon, R. F. Davis y D. E. Newbury. "Self-diffusion of Si-30 in polycrystalline β-SiC". J. Mater. Sci. 15, 2073-2080 (1980).
21. M. H. Hon, R. F. Davis y D. E. Newbury. "Self-diffusion of Silicon-30 in α-SiC single crystals". J. Mater. Sci., 16, 2485 (1981).
22. M. F. Ashby, y R. A. Verrall. "Diffusional accommodated flow and superplasticity". Acta Metall., 21, 149-163 (1973).